\documentstyle[12pt]{article}
\begin{document}
\begin{titlepage}
\begin{flushright}
IASSNS-HEP-95/109\\
\end{flushright}

\vspace{.25cm}
\begin{centering}
{\LARGE{\bf Geodesics around a dislocation}}\\
\bigskip\bigskip
\renewcommand{\thefootnote}{\fnsymbol{footnote}}
Fernando Moraes\footnote{On leave from:\\
Departamento de F\'{\i}sica\\
Universidade Federal de Pernambuco\\
50670-901 Recife, PE, Brazil}\\
{\em School of Natural Sciences\\
Institute for Advanced Study\\
Princeton, NJ 08540\\
U.\ S.\ A.}

\end{centering}
\vspace{1.5cm}
\begin{abstract}
One method of gaining some insight into the motion of particles in a medium
with topological defects (e.g., electrons in a dislocated metal) is to look at
the geodesics of the medium around the defect. In this work the Hamilton-Jacobi
equation for the geodesics in a continuous medium containing a torsional
defect, an edge dislocation, is solved by using perturbation theory to first
order in the Burgers vector.
\end{abstract}

\end{titlepage}
\def\carre{\vbox{\hrule\hbox{\vrule\kern 3pt
\vbox{\kern 3pt\kern 3pt}\kern 3pt\vrule}\hrule}}

\baselineskip = 18pt

Topological defects in solids have been described by Riemann-Cartan geometry
since the early 1950's~\cite{Kro}. More recently, Katanaev and
Volovich~\cite{Kat} have shown the equivalence between three-dimensional
gravity with torsion and the theory of defects in solids. Based in this
formalism, the non-relativistic quantum mechanics of electrons or holes around
disclinations~\cite{Fur1,Fur2} and some properties of the quantum
electromagnetic field also around disclinations~\cite{Mor1,Mor2} have been
studied. Disclinations, defects associated with the breaking of rotational
symmetry, are common in lower dimensional systems like liquid
crystals~\cite{Bow} and an important ingredient in the making of geometrically
frustrated amorphous solids~\cite{Kle}. They do not appear in bulk
crystals,though, due to the very high elastic energy involved in their
formation. Dislocations, associated with translational symmetry breaking, do
appear in ordinary 3D crystals, but the asymmetry the!
 y provoke in the medium makes it difficult to study dynamical properties of
objects moving in their midst. By presenting the geodesics around a
dislocation, this work contributes to the visualization of the dynamics of such
objects and hopefully, will provide a means of obtaining some insight into the
motion of particles and the geometrical optics in dislocated media.

We are concerned here with the geometrical structure of a crystalline medium
that contains a single edge dislocation. An edge dislocation may be
formed~\cite{Kat} by placing a pair of opposing parallel disclinations next to
each other as a dipole (Fig. 1). Each disclination is made by either removing
or inserting a wedge of material.  In the continuum limit, at large distances
from the defect, the medium is described by the metric~\cite{Kat}
\begin{equation}
ds^2=dz^2+\left(1+\frac{m}{2\pi}\frac{2hr \sin \theta -
h^2}{r^2}\right)(dr^2+r^2d\theta^2),
\end{equation}
in cylindrical coordinates. The wedge angles are given by $\pm m$ and the
disclinations are at a distance $h$ apart. The Burgers vector $\vec b=-mh$\^x
is therefore along the x-axis and the defect itself is along the z-axis. Since
the defect preserves translational symmetry along the z-axis, from now on we
will be working in a z=constant surface.

The geodesics may be obtained by requiring that the distance between two given
points in the manifold be a minimum. That is, if $\gamma$ is a curve
parameterized by $t$ joining the points at $t=t_0$ and at $t=t_1$ in the
manifold,
\begin{equation}
ds^2=g_{ij}dx^{i}dx^{j}
\end{equation}
leads to
\begin{equation}
s[\gamma]=\int_{t_0}^{t_1} (g_{ij}\dot{x}^{i} \dot{x}^{j})^{1/2}dt.
\end{equation}
Requiring $s[\gamma]$ to be minimum, determines the path $\gamma$. The geodesic
$\gamma$ is then a solution to the equation
\begin{equation}
\frac{d^{2}x^{i}}{dt^2}+\Gamma^{i}_{jk}\frac{dx^j}{dt}\frac{dx^k}{dt}=0,
\end{equation}
where the Christoffel symbols of the second kind, $\Gamma^{i}_{jk}$, are
\begin{equation}
\Gamma^{i}_{jk}=\frac{1}{2}g^{is}\left(\frac{\partial g_{sj}}{\partial
x^{k}}+\frac{\partial g_{sk}}{\partial x^{j}}-\frac{\partial g_{jk}}{\partial
x^{s}}\right)
\end{equation}
and $g_{ij}g^{ij}=1$.

In the language of Lagrangian mechanics, $s[\gamma]$ is the action functional
and the integrand in equation (3) is the Lagrangian. This procedure, when
applied to the metric (1) leads to a cumbersome pair of coupled differential
equations of second order. We follow instead the Hamilton-Jacobi method
described in~\cite{Baz}. But first we rewrite the z=const. section of metric
(1) in Cartesian coordinates and, since we are interested in effects far away
from the defect (i.e. $r>>h$), neglect the term in $h^{2}/r^{2}$:
\begin{equation}
ds^2=\left(1-\frac{b}{\pi}\frac {y}{x^2+y^2}\right)(dx^2+dy^2).
\end{equation}
Justified by this same argument we will perform below perturbation theory in
powers of $b$.

It is well known~\cite{Mis} that the Newtonian-like action
\begin{equation}
W[\gamma]=\frac{1}{2}\int_{t_0}^{t_1} g_{ij}\dot{x}^{i} \dot{x}^{j}dt
\end{equation}
also leads to the equations of geodesics. Using this action we proceed to write
Hamilton-Jacobi equation
\begin{equation}
\frac{\partial W}{\partial t}+H\left(\frac{\partial W}{\partial
x^{i}}\right)=0,
\end{equation}
where $H=L=\frac{1}{2}g_{ij}\dot{x}^{i} \dot{x}^{j}$ is the Hamiltonian and $L$
is the Lagrangian. Now, writing $H$ in terms of the moments
$p_{i}=\frac{\partial L}{\partial \dot x^{i}}=\frac{\partial W}{\partial
x^{i}}$ we get
\begin{equation}
2\left(1-\frac{b}{\pi}\frac{y}{x^2+y^2}\right)\frac{\partial W}{\partial
t}+\left(\frac{\partial W}{\partial x}\right)^2+\left(\frac{\partial
W}{\partial y}\right)^2=0.
\end{equation}
It follows that
\begin{equation}
W=c_{1}t+\widetilde W(x,y)
\end{equation}
implying
\begin{equation}
\left(\frac{\partial \widetilde W}{\partial x}\right)^2+\left(\frac{\partial
\widetilde W}{\partial
y}\right)^2+2c_{1}\left(1-\frac{b}{\pi}\frac{y}{x^2+y^2}\right)=0.
\end{equation}
Next, we write the perturbation series for $\widetilde W$
\begin{equation}
\widetilde W=\widetilde W_0 +b\widetilde W_1 + b^2 \widetilde W_2 +...
\end{equation}
Substituting (12) into (11) and equating terms of same order in $b$ we find
\begin{equation}
\left(\frac{\partial \widetilde W_0}{\partial x}\right)^2+\left(\frac{\partial
\widetilde W_0}{\partial y}\right)^2+2c_1=0
\end{equation}
and
\begin{equation}
\frac{\partial \widetilde W_0}{\partial x}\frac{\partial \widetilde
W_1}{\partial x}+\frac{\partial \widetilde W_0}{\partial y}\frac{\partial
\widetilde W_1}{\partial y}=\frac{c_1}{\pi}\frac{y}{x^2+y^2}
\end{equation}
up to first order corrections.

Equation (13) is easily solved by the $ansatz$
\begin{equation}
\widetilde W_0=c_{2}x+c_{3}y,
\end{equation}
where the constants $c_{1},c_{2}$ and $c_{3}$ are such that
\begin{equation}
c_1=-\frac{1}{2}(c_{2}^{2}+c_{3}^{2}).
\end{equation}
Now we turn to equation (14) which becomes
\begin{equation}
c_{2}\frac{\partial \widetilde W_1}{\partial x}+c_{3}\frac{\partial \widetilde
W_1}{\partial y}=\frac{c_1}{\pi}\frac{y}{x^2+y^2}.
\end{equation}
This equation is solved by the method of characteristics~\cite{Zwi}:
first differentiate $W_1$ with respect to the auxiliary variable u,
\begin{equation}
\frac{d\widetilde W_1}{du}=\left(\frac{\partial x}{\partial
u}\right)\left(\frac{\partial \widetilde W_1}{\partial
x}\right)+\left(\frac{\partial y}{\partial u}\right)\left(\frac{\partial
\widetilde W_1}{\partial y}\right),
\end{equation}
then by comparing (17) and (18) make the identifications
\begin{equation}
\frac{d\widetilde W_1}{du}=\frac{c_1}{\pi}\frac{y}{x^2+y^2}, \;
\frac{\partial x}{\partial u}=c_2, \;
\frac{\partial y}{\partial u}=c_3.
\end{equation}
It follows that
\begin{equation}
x=c_{2}u+t_{1},\;and\;y=c_{3}u,
\end{equation}
where $t_1$ is another auxiliary variable.
Now we substitute (20) in the first equation in (19) obtaining
\begin{equation}
\frac{d\widetilde
W_1}{du}=\frac{c_{1}c_{3}}{\pi}\frac{u}{(c_{2}u+t_1)^2+(c_{3}u)^2}.
\end{equation}
Integration with respect to $u$ of the above equation, followed by substitution
of $u=y/c_3$ and $t_{1}=x-\frac{c_2}{c_3}y$ (from (20)) and use of (16) yields
\begin{equation}
W \approx c_{1}t+ \widetilde W_{0}+b\widetilde W_{1}
=-\frac{1}{2}(c_{2}^{2}+c_{3}^{2})t+c_{2}x+c_{3}y-\frac{c_{3}b}{4\pi}\ln
(x^2+y^2)+\frac{c_{2}b}{2\pi}\arctan \left(
\frac{c_{2}x+c_{3}y}{c_{3}x-c_{2}y}\right).
\end{equation}
The corresponding solution (up to a constant $A$) of the equations of motion is
obtained by differentiation with respect to $c_2$ and $c_3$:
\begin{equation}
-c_{2}t+x+\frac{b}{2\pi}\arctan \left(
\frac{c_{2}x+c_{3}y}{c_{3}x-c_{2}y}\right)+\frac{b}{2\pi}\left(
\frac{c_{2}c_{3}}{c_{2}^{2}+c_{3}^{2}}\right)=0
\end{equation}
and
\begin{equation}
-c_{3}t+y-\frac{b}{4\pi}\ln (x^2+y^2)-\frac{b}{2\pi}\left(
\frac{c_{2}^{2}}{c_{2}^{2}+c_{3}^{2}}\right)=0.
\end{equation}
Inserting the missing constant, eliminating $t$ between the above equations and
making the substitution $B=c_{3}/c_{2}$, we obtain
\begin{equation}
y=A+Bx+\frac{b}{2\pi}\left[ \frac{1}{2} \ln(x^2+y^2)+B\arctan \left(
\frac{x+By}{Bx-y}\right)+1\right],
\end{equation}
which gives the correct limit in the absence of the defect ($b=0$). A first
iteration of this equation (still keeping to first order in $b$) allows us to
plot the geodesics, which are shown in Figs. 2-4. Each figure shows a bundle of
geodesics parallel to each other at infinity, approaching and being deflected
by the dislocation located at the origin of the plots.

Due to the particular choice of coordinates, plotting geodesics that
asymptotically are of the kind x=const. (i.e. vertical), is tricky. To do so,
we first divide equation (25) by $B$ and then take the limit
$B\rightarrow\infty$. We are left with
\begin{equation}
C=x+\frac{b}{2\pi}\arctan \left(\frac{y}{x}\right),
\end{equation}
where $C$ is constant. As before, this gives the correct vertical geodesics in
the limit $b=0$.
Inversion of the above equation gives
\begin{equation}
y=x\tan \left[ \frac{2\pi}{b}(C-x) \right].
\end{equation}
Because of the periodicity of $\tan \theta$  we need only to choose a
convenient range for $\theta=\frac{2\pi}{b}(C-x)$ in order to plot the
geodesics. Since our approach is only valid away from the origin, the range
$-\pi/2<\theta<\pi/2$ is excluded. We use then the ranges
$-3\pi/2<\theta<-\pi/2$ and $\pi/2<\theta<3\pi/2$. The result is shown in Fig.
5.

In conclusion, the solution to the equation of geodesics in a medium containing
one edge dislocation has been found by using perturbation theory to first order
in the Burgers vector. A single dislocation produces a dilatation that breaks
the translational symmetry of the crystalline lattice causing a deformation of
the geodesics around the defect as it is clear from Figs. 2-5. This gives an
intuitive picture to what happens to (classical) electrons moving in a
dislocated metal or to the propagation of rays in a dislocated continuous
medium, for example.

\noindent
{\bf Acknowledgment} This work was partially supported by CNPq.

\noindent
{\Large {\bf Figure Captions}}
\\
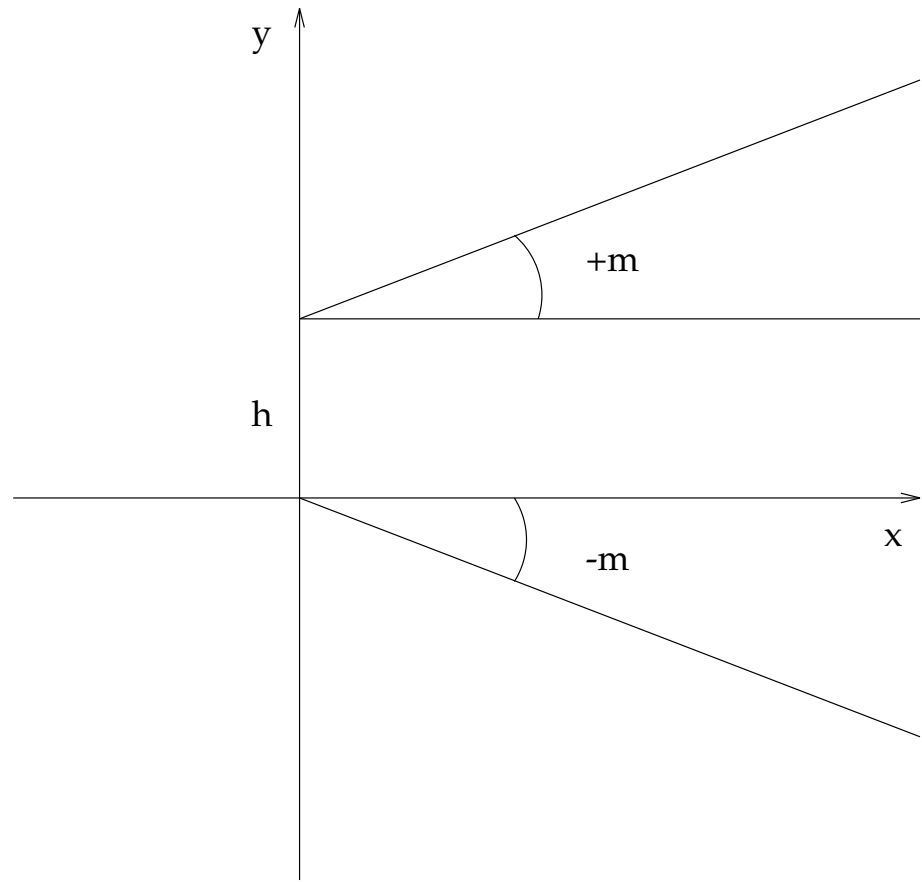
Figure 1: Disclination dipole making up an edge dislocation.
\\
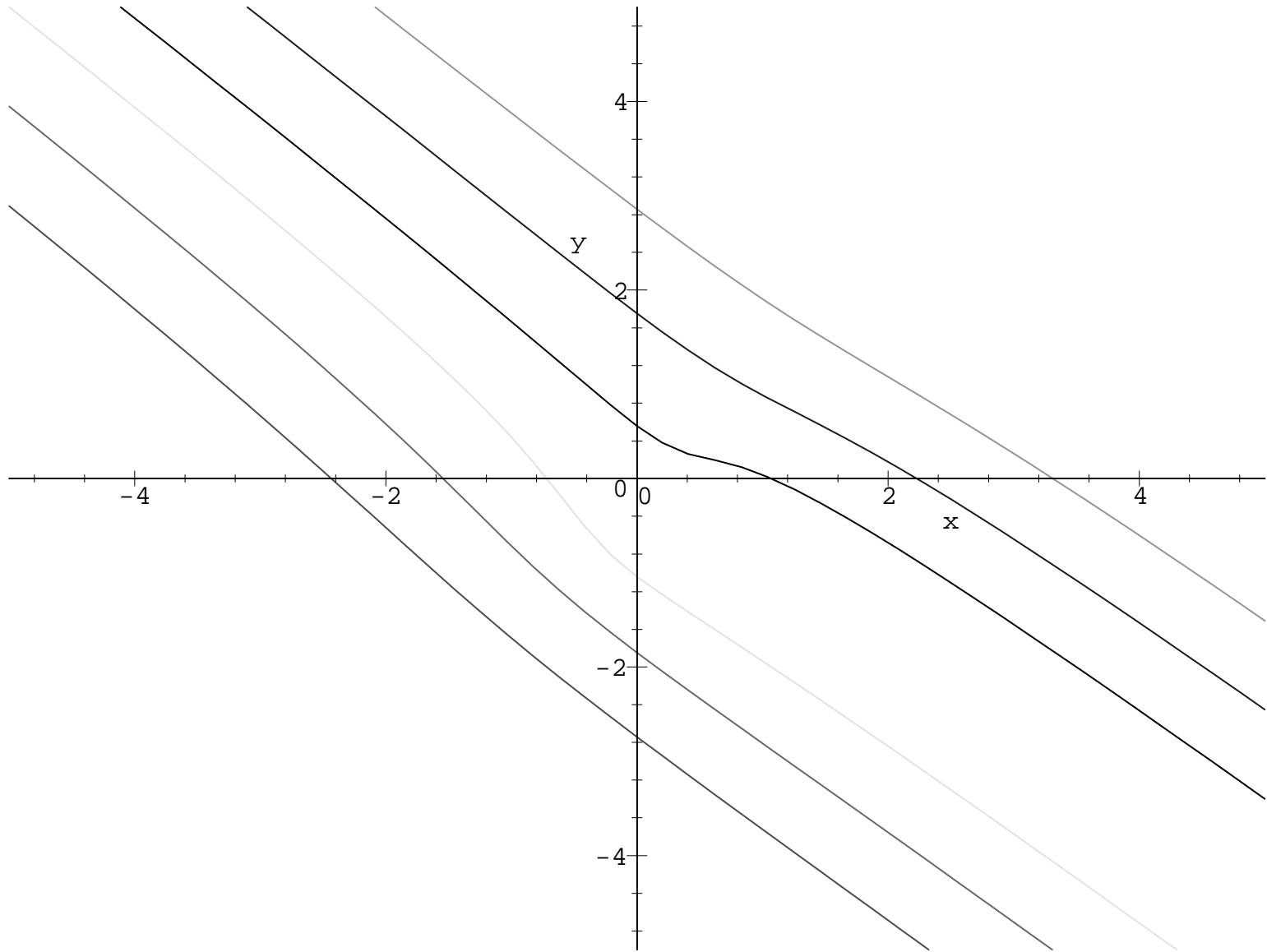
Figure 2: Geodesics around edge dislocation, $B=-1$.
\\
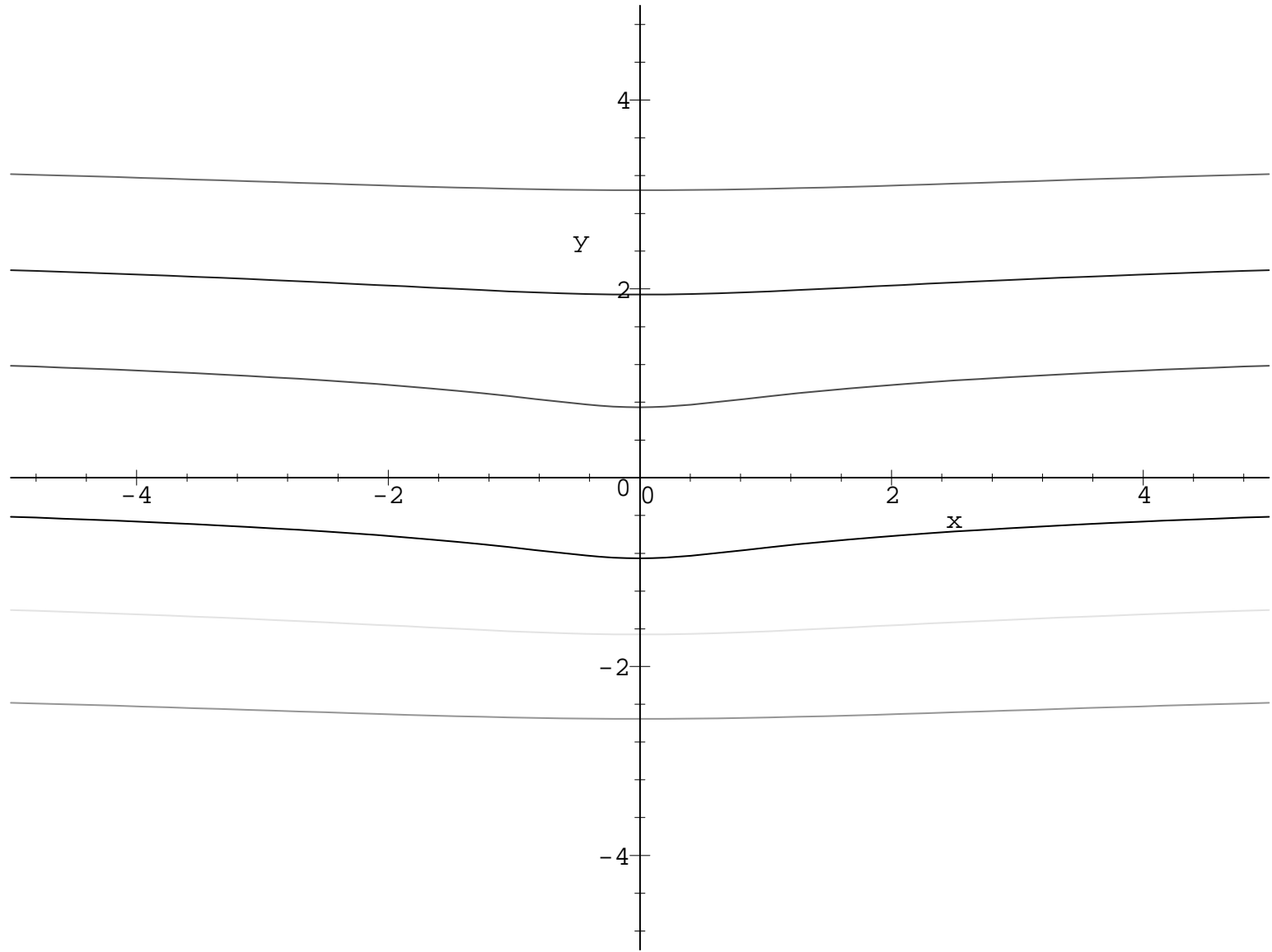
Figure 3: Geodesics around edge dislocation, $B=0$.
\\
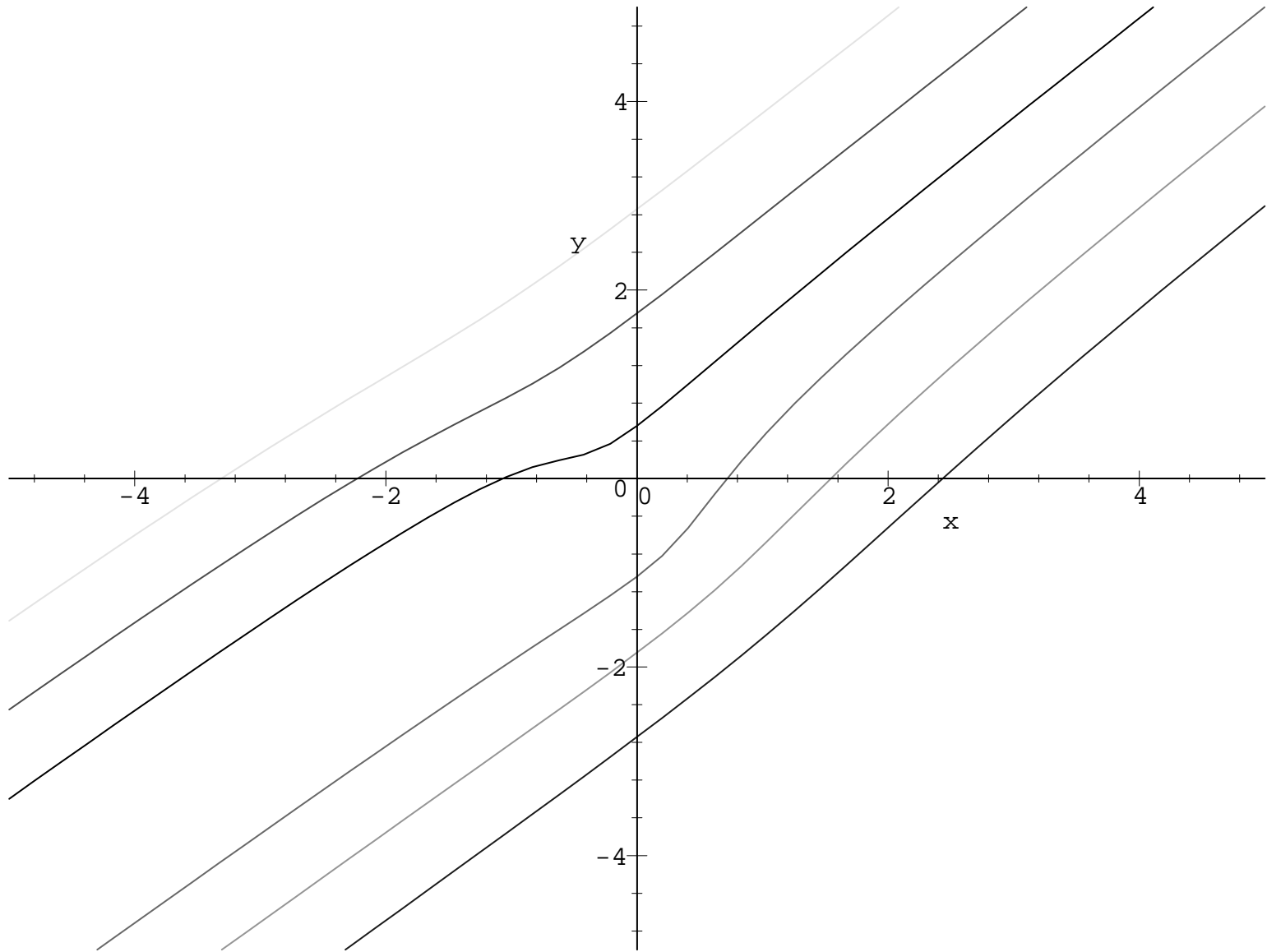
Figure 4: Geodesics around edge dislocation, $B=1$.
\\
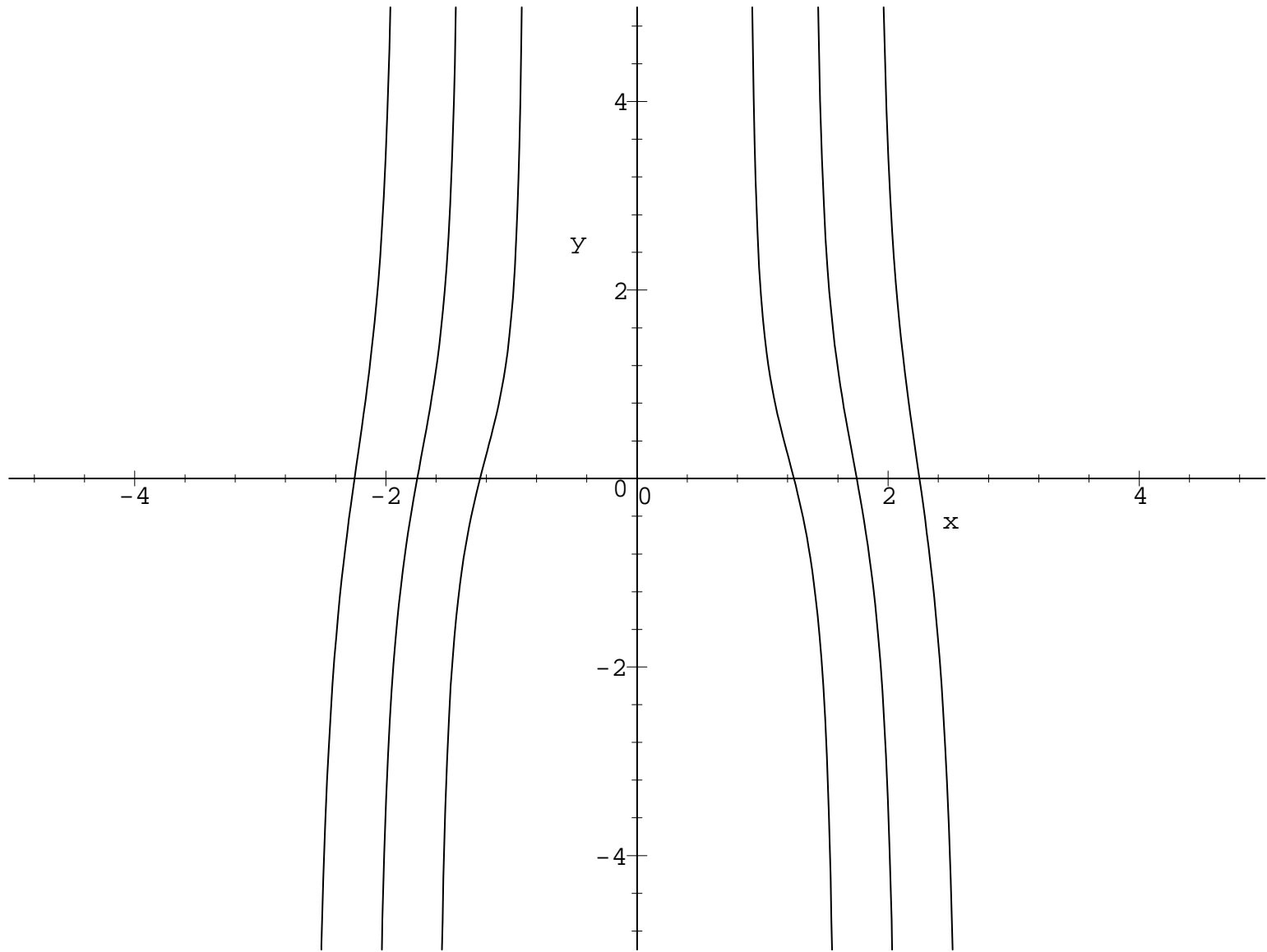
Figure 5: Geodesics around edge dislocation, $B=\infty$.
\end{document}